\documentclass[fp,twocolumn]{jpsj3}
\usepackage{amsmath}
\usepackage{amssymb}
\usepackage{bm}
\usepackage{graphicx}
\usepackage{times}
\usepackage{txfonts}
\usepackage{booktabs}
\usepackage[usenames]{color}

\title{
First-principles Study of the RKKY Interaction and the Quadrupole Order in the Pr 1-2-20 systems Pr$T_{2}$Al$_{20}$ ($T$=Ti, V)
}
\author{
Yuto {\sc Iizuka}$^1$, 
Takemi {\sc Yamada}$^2$,
Katsurou {\sc Hanzawa}$^2$, and 
Yoshiaki {\sc \=Ono}$^1$\thanks{E-mail address: y.ono@phys.sc.niigata-u.ac.jp}
}
\inst{
$^1$Department of Physics, Niigata University, Niigata, 950-2181, Japan\\
$^2$Department of Physics, Faculty of Science and Technology, Tokyo University of Science, Chiba, 278-8510, Japan
}
\abst{
Electronic states and quadrupole orders in the Pr 1-2-20 systems Pr$T_{2}$Al$_{20}$ ($T$=Ti, V) are investigated on the basis of the first-principles calculations. The effective 196 orbital model is derived to reproduce the first-principles electronic structures of La$T_{2}$Al$_{20}$ ($T$=Ti, V) without contribution from the Pr 4$f$ electrons which are considered to be well localized and is employed to calculate the Ruderman-Kittel-Kasuya-Yosida (RKKY) interactions between quadrupole and octupole moments of the Pr ions. Within the random phase approximation for the RKKY Hamiltonian, the most divergent susceptibility is found to be the quadrupole one for the wave vector ${\bm{Q}}=(0,0,0)$ in the case of PrTi$_{2}$Al$_{20}$ while that for  ${\bm{Q}}=(\pi/a,0,\pi/a)$  in the case of PrV$_{2}$Al$_{20}$  as consistent with experimental observations in the both cases which exhibit the ferro-quadrupole (FQ) and the antiferro-quadrupole (AFQ) orders, respectively. We also discuss the ordered states using the mean-field approximation and find that, in the case of PrTi$_{2}$Al$_{20}$, the 1st-order phase transition to the $O_2^0$ FQ order with a tiny discontinuity takes place as predicted by the Landau theory. In the case of PrV$_{2}$Al$_{20}$, the system exhibits two distinct $O_2^2$ AFQ orders, AFQ-I and AFQ-II, and shows subsequent two phase transitions, the 2nd-order one from normal to AFQ-I and the 1st-order one from AFQ-I to AFQ-II, that may be responsible for the double transitions observed by specific heat measurements. 
}

\begin{document}
\maketitle

\section{Introduction}
\label{section1}

Recently, the Pr 1-2-20 systems Pr$T_{2}X_{20}$ have attracted much attention as they show various interesting phenomena including quadrupole orders, superconductivity and non Fermi liquid behaviors\cite{Onimaru2016}. The crystal structure is the cubic CeCr$_{2}$Al$_{20}$-type with the space group Fd$\overline{3}$m\cite{Niemann1995}, where the Pr atoms are encapsulated in the Frank-Kasper cages formed by 16 $X$ atoms\cite{Frank1958} and are under $T_{\rm d}$-symmetry. The  crystalline electric field (CEF) ground states of the Pr ions with the $4f^2$ configuration are the non-Kramers doublets $\Gamma_{3}$, where we can elucidate the effects of the electric quadrupoles without taking into account of the magnetic moments. 

In fact, PrTi$_{2}$Al$_{20}$ exhibits the ferro-quadrupole (FQ) order below the transition temperature $T_{\rm Q}=2.0$ K\cite{Koseki2011,Sakai2011} at the ambient pressure, and PrV$_{2}$Al$_{20}$, PrIr$_{2}$Zn$_{20}$ and  PrRh$_{2}$Zn$_{20}$ exhibit the antiferro-quadrupole (AFQ) order below $T_{\rm Q}=0.6$, 0.11 and 0.06 K, respectively\cite{Ishii2011,Ishii2013,Onimaru2010,Onimaru2012}. Among them, the ordered quadrupole moments have been determined as $O_2^0$ for the FQ order in PrTi$_{2}$Al$_{20}$\cite{Sato2012,Taniguchi2016} while $O_2^2$ for the AFQ order in PrIr$_{2}$Zn$_{20}$\cite{Iwasa2017}. In addition, high-quality single crystals in PrV$_{2}$Al$_{20}$ exhibit remarkable double transitions at $T_{\rm Q}=0.75$ K and $T^* = 0.65$ K\cite{Tsujimoto2015} below which another quadrupole and/or octupole order is expected to be realized. More recently, the FQ order of PrTi$_{2}$Al$_{20}$ has been found to show magnetic-field-induced 1st-order phase transitions\cite{Taniguchi2019,Kittaka2020} which are well reproduced by the Landau theory based on the mean-field approximation\cite{Hattori2014}. 

As is well known, the details of the energy band dispersions near the Fermi level responsible for the Fermi surfaces are important to discuss what kinds of ordered states including quadrupole one take place. In the 1-2-20 systems, the de Haas-van Alphen (dHvA) experiments revealed that the Fermi surfaces for PrTi$_{2}$Al$_{20}$ and PrIr$_{2}$Zn$_{20}$ are well accounted for by the first-principles band calculations for LaTi$_{2}$Al$_{20}$ and LaIr$_{2}$Zn$_{20}$ without contribution from 4$f$ electrons  of Pr ions, respectively\cite{Nagashima2014,Matsushita2011}, which is interpreted as indicating that the $4f$ electrons are sufficiently localized. Then, the quadrupole order of the localized $4f$ electrons is considered due to the Ruderman-Kittel-Kasuya-Yosida (RKKY) interaction\cite{Ruderman1954,Kasuya1956,Yosida1957} between the quadrupole moments of Pr ions mediated by the conduction ($c$) electrons, where details of the electronic structure of the $c$ electrons are important to determine the RKKY interaction\cite{Hanzawa2015,Yamada2019}. 

In the previous paper\cite{Iizuka2020}, we have calculated the RKKY interaction between the quadrupole moments of the Pr ions in Pr$T_{2}$Al$_{20}$ ($T$=Ti, V) on the basis of the realistic tight binding models for the $c$ electrons extracted from the first-principles band calculation and have found that the wave vector of the expected quadrupole order is ${\bm{Q}}=\left(0,0,0\right)$ in the case of PrTi$_{2}$Al$_{20}$ while it is ${\bm{Q}}=\left(\pi/a,0,\pi/a\right)$ in the case of PrV$_{2}$Al$_{20}$ as consistent with experimental observations in PrTi$_{2}$Al$_{20}$ and PrV$_{2}$Al$_{20}$ which exhibit FQ and AFQ orders, respectively. The effects of the octupole moments and the ordered states below $T_{\rm Q}$, however, were not considered there\cite{Iizuka2020}. The present paper is a straight-forward extension of our previous work\cite{Iizuka2020} to include the RKKY interaction between the octupole moments in addition to the quadrupole ones and to explicitly discuss the ordered states below $T_{\rm Q}$, that is important for comparison with the experiments.

\section{Model and Formulation}
\label{section2}

\begin{figure}[t]
\centering
\vspace{+0.3cm} 
\includegraphics[width=8.0cm]{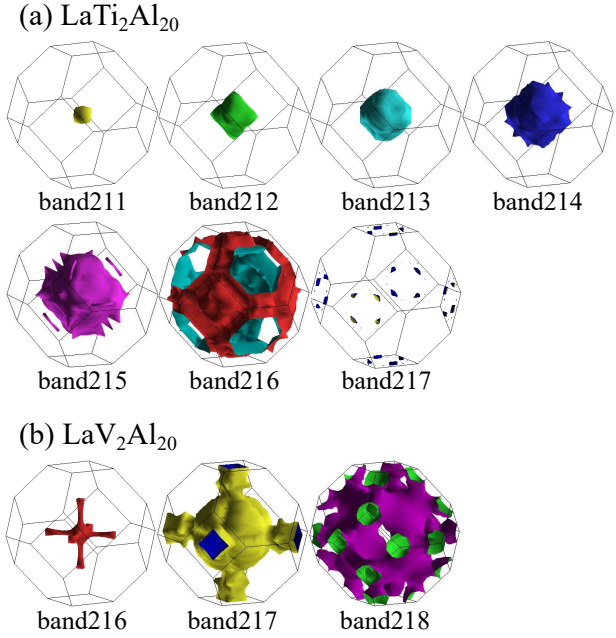}
\caption{(color online) 
Fermi surfaces of (a) LaTi$_{2}$Al$_{20}$ and (b) LaV$_{2}$Al$_{20}$ from the first-principles band calculations, where band211-215 in (a) and band216 in (b) are hole Fermi surfaces and the others are electron ones.}
\label{Figure1}
\end{figure}

First, we perform the first-principles band calculations for  La$T_{2}$Al$_{20}$ ($T$=Ti, V) on the basis of the density-functional theory (DFT) with the generalized gradient approximation (GGA) by using WIEN2k code\cite{Blaha2001}, where $17\times17\times17$ ${\bm{k}}$-points, the muffin-tin radius $R_{\rm MT} = 2.5$ (2.3) a.u. for La and $T$ (Al) and the plane-wave cuttoff $K_{\rm max}=3.2$ (a.u.)$^{-1}$ are used. As for the lattice parameters, we employ the experimentally determined values of Pr$T_{2}$Al$_{20}$ ($T$=Ti, V)\cite{Kangas2012} instead of La$T_{2}$Al$_{20}$ in order to discuss the former electronic states in the 4$f$ electron localized regime. In addition, we use the GGA+$U$ method with an artificially large value of $U=60$ eV to exclude the La-4$f$ components near the Fermi level. Figures 1 (a) and (b) show the obtained Fermi surfaces of La$T_{2}$Al$_{20}$ ($T$=Ti, V) which are in good agreements with those in the previous studies\cite{Nagashima2014,Swatek2018}. 

Next, we derive the tight binding models for the conduction electrons so as to reproduce the first-principles electronic structures near the Fermi level for La$T_{2}$Al$_{20}$ ($T$=Ti, V) as shown in Figs. 2 (a) and (b), respectively, by using the maximally localized Wannier functions\cite{Kunes2010}  which consist of 196 orbitals as a primitive unit cell includes La$_2$$T_{4}$Al$_{40}$: La-$d$ (5 orbitals $\times$ 2 sites), La-$s$ (1 orbital $\times$ 2 sites), $T$-$d$  (5 orbitals $\times$ 4 sites), $T$-$s$  (1 orbital $\times$ 4 sites), Al-$p$ (3 orbitals $\times$ 40 sites) and Al-$s$ (1 orbital $\times$ 40 sites) in the conventional unit cell\cite{Iizuka2020}. When Ti is substituted by V in  Pr$T_2$Al$_{20}$ ($T$=Ti, V), the number of electron increases and then the Fermi level shifts upward as shown in Figs. 2 (a) and (b) resulting in the disappear of the small hole FSs of the band211-214 centered at the $\Gamma$ point (see Fig. 1 (a)) which are crucial for a ferro-component of the susceptibility in the case of PrTi$_{2}$Al$_{20}$ as mentioned later.

\begin{figure}[t]
\centering
\vspace{+0.3cm}
\includegraphics[width=8.48cm]{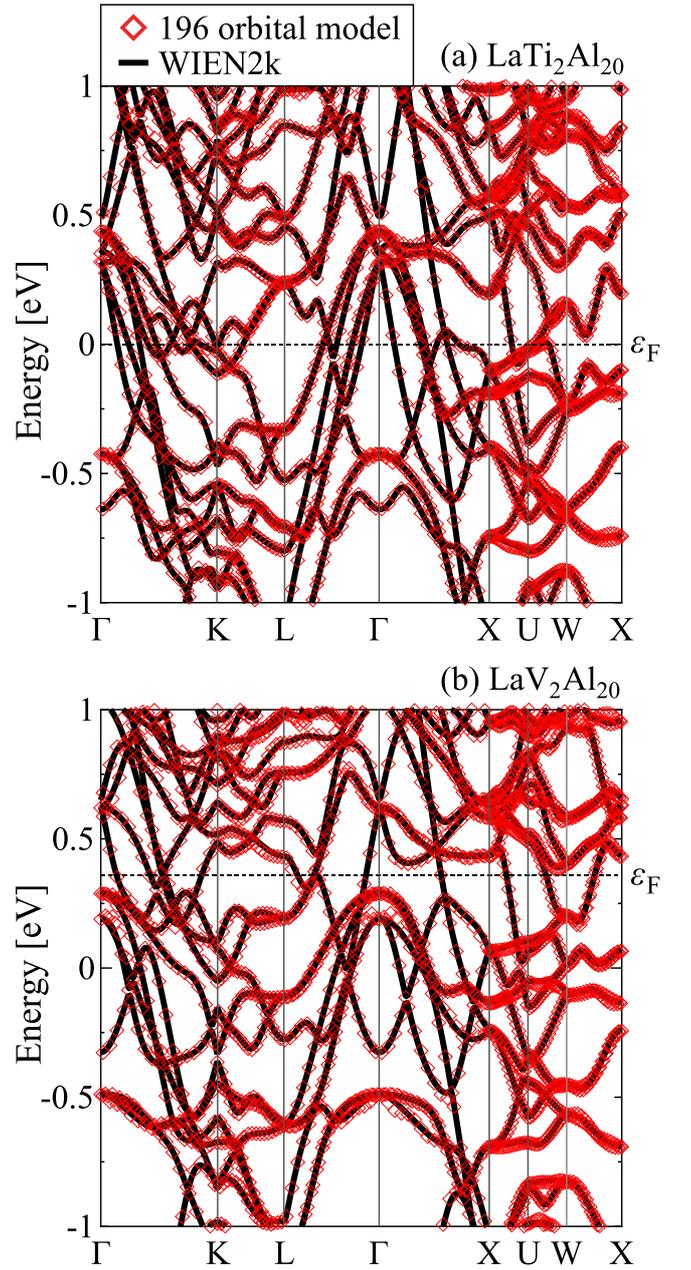}
\caption{(color online) 
Energy band dispersions near the Fermi level $\varepsilon_{\rm{F}}$ (black dashed line) of LaTi$_{2}$Al$_{20}$ (a) and LaV$_{2}$Al$_{20}$ (b)  from the 196 orbital models (red diamonds) together with those from the first-principles band calculations using WIEN2k (black solid lines) as functions of the wave vector along the symmetric lines through the following sysmmetric points: $\Gamma:(0,0,0)$, ${\rm K}:(3\pi/4a,3\pi/4a,3\pi/2a)$, ${\rm L}:(\pi/a,\pi/a,\pi/a)$, ${\rm X}:(\pi/a,0,\pi/a)$, ${\rm U}:(5\pi/4a,\pi/2a,5\pi/4a)$ and ${\rm W}:(\pi/a,\pi/2a,3\pi/2a)$, where the energies are plotted relative to the Fermi level of LaTi$_{2}$Al$_{20}$ in the both (a) and (b) and $\varepsilon_{\rm{F}}=0.38$eV in (b) indicates the relative shift of the Fermi level of LaV$_{2}$Al$_{20}$.
}
\label{Figure2}
\end{figure}

The RKKY interaction between the localized $f$ electrons is the indirect interaction mediated by the $c$ electrons via the hybridization between the $c$ and $f$ electrons\cite{Ruderman1954,Kasuya1956,Yosida1957}. 
As the Pr $5d$ electrons give the dominant contribution to the susceptibility responsible for the RKKY interaction, we consider the $c$-$f$ hybridization only between the $5d$ and $4f$ on the same Pr atom which becomes finite for the CEF Hamiltonian ${\cal H}_{\rm CEF}$ without inversion symmetry\cite{Hanzawa2011} and is explicitly given by
\begin{align}
V_{mm'\ell \sigma }^{\left( n \right)} &= \bigl\langle {{f^n}m } \bigr|{{\cal H}_{\rm CEF}} \bigl| {{f^{n - 1}}m'} \bigr\rangle | {\ell \sigma } \rangle,
\end{align}
where $\left| {{f^{n}}m} \right\rangle$ is a state of the Pr $4f^n$ configuration labeled by $m$ and $\left| {\ell \sigma } \right\rangle$ is a Pr $5d$ state with the orbital $\ell$ and spin $\sigma$, where the $t_{2g}$ orbital exclusively hybridizes with the $f$ orbital in the case with the $T_{\rm d}$-symmetric CEF Hamiltonian. The explicit dependence of $V_{mm'\ell \sigma }^{(n)}$  on $m, m', \ell$ and $\sigma$ is determined by using the Clebsch-Gordan coefficients\cite{Hanzawa2015} (not shown) except for the absolute value proportional to the strength of the CEF potential $V_{\rm CEF}$ which is an unknown parameter here.

In the non-Kramers doublet ground states $\Gamma _{3}$ of Pr $4f^2$, the second-order perturbation with respect to $V_{mm'\ell \sigma }^{\left( n \right)}$ yields the Kondo exchange interaction Hamiltonian
\begin{align}
\label{Hex}
{H}_{\rm ex} = \sum_{i} \sum_{\mu  = 1,2} \sum_{\alpha=x,y,z} \sum_{\ell \ell ' = t_{2g}} \sum_{\sigma \sigma '=\uparrow,\downarrow} K_{\ell \ell '\sigma \sigma' }^\alpha \hat X_\alpha ^\mu (i) d_{i\nu \ell \sigma }^\dag d_{i\nu \ell '\sigma '},
\end{align}
where $d_{i\nu \ell\sigma }^\dag$ is the creation operator for a Pr $5d$-$t_{2g}$ electron with orbital $\ell$ and  spin $\sigma$ at Pr site $\mu$ of the $i$-th unit cell, and $\hat X_{\alpha}$ is the multipole operator explicitly defined as follows: the quadrupoles $O_2^0$ for $\alpha=z$ and  $O_2^2$ for $\alpha=x$ and the octupole $T_{xyz}$ for $\alpha=y$. In eq. (\ref{Hex}), the Kondo coupling is given by
\begin{align}
\label{Kondo}
K_{\ell \ell '\sigma \sigma '}^\alpha 
 &= \frac12 \sum_{mm'=\Gamma_3}  \sum_{M}  
\left\langle {{f^2}m' } \right|\hat X_\alpha \left| {{f^{2}}m} \right\rangle
\nonumber \\ 
 &\times \left( D_{+}V_{Mm\ell \sigma }^{(3)*} V_{M m'\ell '\sigma '}^{(3)}
 + D_{-}V_{m'M \ell \sigma }^{(2)*} V_{mM \ell '\sigma '}^{(2)}  \right) 
\end{align}
with 
\begin{align}
\label{D+}
D_{+}&=1/(E_M^{\left( 3 \right)} - E_{m}^{\left( 2 \right)} - \varepsilon _{\rm F}), 
\\
\label{D-}
D_{-}&=1/(\varepsilon _{\rm F} + E_M^{\left( 1 \right)} - E_{m}^{\left( 2 \right)}),  
\end{align}
where
$E_{m}^{(2)}$ and $E_M^{1(3)}$ are the energies for the state $m$ of the Pr $4f^2$ with the non-Kramers doublets $\Gamma _{3}$ and the state $M$ of the Pr $4f^{1(3)}$ with $J=5/2\ (9/2)$, and $\varepsilon _{\rm F}$ is the Fermi level around which the energy bands largely consist of Pr $5d$-$t_{2g}$ orbitals exist. Among the energies, $E_M^{(1)}$ and $\varepsilon _{\rm F}$ can be extracted from the first-principles band calculation, but $E_{m}^{(2)}$ and $E_M^{(3)}$ including the effects of the Coulomb interaction between the Pr $4f$ electrons are not explicitly determined here.

From the second-order perturbation with respect to $H_{\rm ex}$ in eq. (\ref{Hex}), we obtain the RKKY  Hamiltonian between the quadrupoles  $\hat X_{z} =\hat O_2^0$ and  $\hat X_{x} =\hat O_2^2$ and the octupole $\hat X_{y} =\hat T_{xyz}$ at the Pr site $\mu(\nu)$ of the $i(j)$-th unit cell as
\begin{align}
\label{Hrkky_multipoles}
        H_{\rm{RKKY}} =  - \sum\limits_{\langle i,j \rangle}
 \sum\limits_{\mu \nu  = 1,2} \sum\limits_{\alpha \beta  = x,y,z} 
 J_{\alpha \beta }^{\mu \nu }(i,j) \hat X_\alpha ^\mu (i) \hat X_\beta ^\nu (j),
\end{align}
with the RKKY interaction given by
\begin{align}
\label{Jrkky}
        J_{\alpha \beta }^{\mu \nu } (i,j) = \frac1N \sum_{\bm{q}}
 \sum\limits_{\{ \ell \}}  \sum\limits_{\sigma \sigma'}
 K_{\ell_1 \ell_2 \sigma \sigma'}^\alpha 
 K_{ \ell_3 \ell_4 \sigma \sigma'}^{\beta *}
 \chi_{\ell_1 \ell_2 \ell_3 \ell_4}^{\mu \nu}(\bm{q})
 e^{i\bm{q} \cdot (\bm{r}_i-\bm{r}_j)},
\end{align}
where $\bm{q}$ is  the wave vector, $\bm{r}_i$ is the position of the $i$-th unit cell and $N$ is the total number of the unit cell. In eq. (\ref{Jrkky}), the susceptibility for the $c$ electrons is given by 
\begin{align}
\label{chi}
        \chi _{{\ell _1}{\ell _2}{\ell _3}{\ell _4}}^{\mu \nu }\left( {\bm{q}} \right) = \frac{1}{N}&\sum\limits_{{\bm{k}}ss'}^{} {u_{{\ell _1}s}^{\mu *}\left( {\bm{k}} \right)u_{{\ell _2}s'}^{\mu *}\left( {{\bm{k}} + {\bm{q}}} \right)u_{{\ell _3}s}^{\nu *}\left( {\bm{k}} \right)u_{{\ell _4}s'}^{\nu *}\left( {{\bm{k}} + {\bm{q}}} \right)} \notag\\
        &\times \frac{{f( {{\varepsilon _{s'}}( {{\bm{k}} + {\bm{q}}} )  } ) 
- f( {{\varepsilon _s}( {\bm{k}} )  } )}}{{{\varepsilon _s}\left( {\bm{k}} \right) - {\varepsilon _{s'}}\left( {{\bm{k}} + {\bm{q}}} \right)}},
\end{align}
where $\varepsilon_s (\bm{k})$ is the energy for the $c$ electrons with the wave vector $\bm{k}$ and the band $s$ and $u_{\ell s}\left(\bm{k}\right)$ is the corresponding eigen vector of the component of the orbital $\ell$, and $f(\varepsilon)$ is the Fermi distribution function.

\section{Results}
\label{section3}

Substituting  $\varepsilon _{s}(\bm{k})$ and  $u_{\ell s}(\bm{k})$ obtained from the 196  orbital models into eq. (\ref{chi}), we calculate the RKKY interaction eq. (\ref{Jrkky}) with eqs. (\ref{Kondo}) and (\ref{chi}), where we set temperatures $T=0.01-0.05$ eV around and below which $\chi _{{\ell _1}{\ell _2}{\ell _3}{\ell _4}}^{\mu \nu }(\bm{q})$ is found to be almost independent of $T$. We note that the 196  orbital models are necessary to obtain $u_{\ell s}(\bm{k})$, although $\varepsilon _{s}(\bm{k})$ can be obtained directly from the first-principles band calculations. In order to evaluate the Kondo coupling in eq. (\ref{Kondo}), there are two individual unknown parameters $D_+$ and $D_-$ in which the effect of another unknown parameter, the strength of $V_{\rm CEF}$, can be included. Hereafter, we set the value of $D_-$ with keeping the ratio $D_+/D_- =0.6 \ (1.0)$\cite{ratio} so as to reproduce the experimentally observed transition temperature $T_{\rm Q}$ of the FQ (AFQ) for PrTi$_2$Al$_{20}$ (PrV$_2$Al$_{20}$) within the mean-field approximation for the RKKY Hamiltonian eq. (\ref{Hrkky_multipoles}).

\begin{figure}[t]
\centering
\vspace{+0.3cm}
\includegraphics[width=8.5cm]{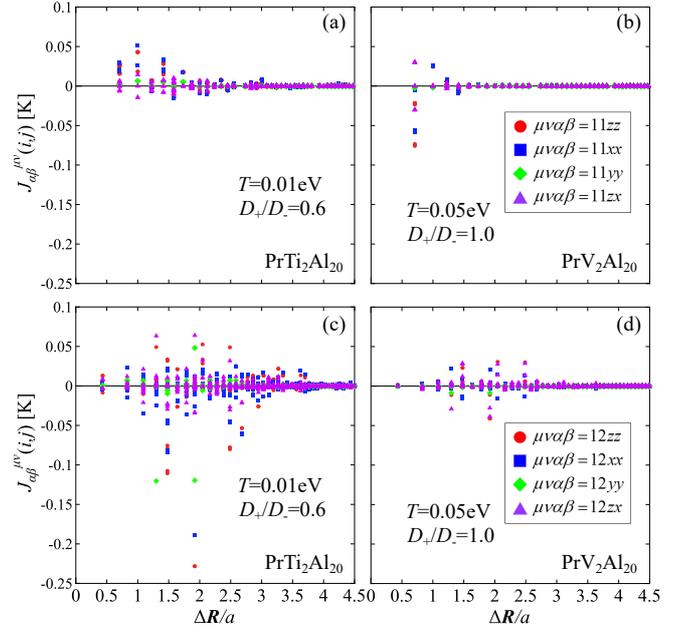}
\caption{(color online) 
Spatial dependence of the RKKY interactions of the diagonal components for $O_2^0$ $(\alpha\beta=zz)$, $O_2^2$ $(\alpha\beta=xx)$ and $T_{xyz}$ $(\alpha\beta=yy)$ and those of the off-diagonal component of $O_2^0$-$O_2^2$ $(\alpha\beta=zx)$ between the same Pr sites $\mu=\nu=1$ (a)(b), and those between the different Pr sites $\mu=1$ and $\nu=2$ (c)(d) in the cases of PrTi$_{2}$Al$_{20}$ (a)(c) and  PrV$_{2}$Al$_{20}$ (b)(d), respectively, where $\Delta {\bm R} \equiv |\bm{R}_i^\mu-\bm{R}_j^\nu|$ with the position $\bm{R}_i^\mu$ of the Pr ion at site $\mu$ of the $i$-th unit cell and $a=14.725$\AA \ (14.567\AA) is the lattice constant for PrTi$_{2}$Al$_{20}$ (PrV$_{2}$Al$_{20}$).}
\label{Figure3}
\end{figure}

Figures 3 (a)-(d) show the  RKKY interactions $J_{\alpha \beta }^{\mu \nu } (i,j)$ as functions of the relative disance $\Delta {\bm R}$ between the Pr ions in the cases of PrTi$_2$Al$_{20}$ ((a) and (c)) and PrV$_2$Al$_{20}$ ((b) and (d)), respectively, where the parameters $T$ and $D_+/D_-$ are set as mentioned in the preceding paragraph.  We observe that $J_{\alpha \beta }^{\mu \nu } (i,j)$ exhibit oscillatory decreases with $\Delta {\bm R}$ more than four unit cells (62th nearest neighbor sites). As shown in Fig. 3 (a) and (b), the diagonal components for the quadrupoles $O_2^0$ and $O_2^2$ between the nearest neighbor sites are positive (negative) responsible for the same (opposite) sign of the quadrupoles yielding the FQ (AFQ) order in the case of PrTi$_2$Al$_{20}$ (PrV$_2$Al$_{20}$). We note that, not only the diagonal components for the quadrupoles $O_2^0$ and $O_2^2$ but also the diagonal component for the octupole $T_{xyz}$ and the off-diagonal components between $O_2^0$ and $O_2^2$ have significant values as shown in Figs. 3 (c) and (d)\cite{maximum}.

\begin{figure}[t]
\centering
\vspace{+0.3cm}
\includegraphics[width=8.6cm]{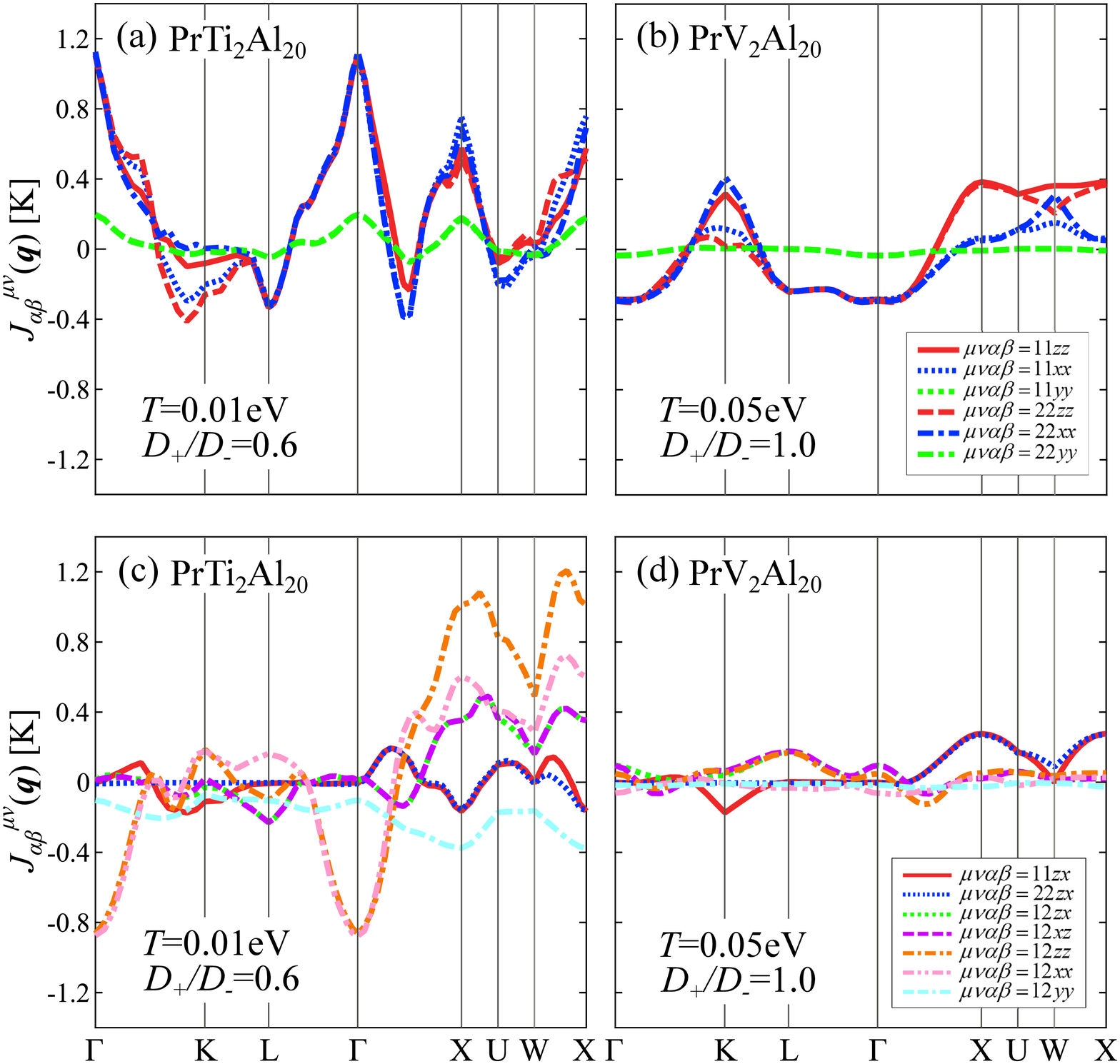}
\caption{(color online) 
Fourier transforms of the RKKY interactions of diagonal components (a)(b) and off-diagonal components with respect to the multipoles and/or the Pr sites (c)(d) in the cases of PrTi$_{2}$Al$_{20}$ (a)(c) and PrV$_{2}$Al$_{20}$ (b)(d), respectively, as functions of the wave vector along the symmetric lines through the sysmmetric points as the same as in Fig. 2. We note that the off-diagonal components of $J_{\alpha \beta }^{\mu \nu } (\bm{q})$ are generally complex and only the real parts are plotted in (c) and (d). 
}
\label{Figure4}
\end{figure}

To discuss the expected multipole orders and the wave vectors $\bm{q}$ of the ordered states, we perform the Fourier transformation of the RKKY interaction. Figures 4 (a)-(b) show the $\bm{q}$-dependence of the Fourier transformed RKKY interactions $J_{\alpha \beta }^{\mu \nu } (\bm{q})$ in the cases of PrTi$_{2}$Al$_{20}$ ((a) and (c)) and PrV$_2$Al$_{20}$ ((b) and (d)), respectively, where the parameters $T$ and $D_+/D_-$ are the same as those in Figs.3 (a)-(d). We find that  the diagonal components for the quadrupoles has a maximum at ${\bm{Q}}=(0,0,0)$ in the case of PrTi$_{2}$Al$_{20}$ (see Fig. 4 (a)) and at ${\bm{Q}}=(\pi/a,0,\pi/a)$ in the case of PrV$_{2}$Al$_{20}$ (see Fig. 4 (b)), while the diagonal component for the  octupole is smaller than the quadrupole components. Within the RPA, these results indicate that the FQ and the AFQ orders take place in the cases of PrTi$_{2}$Al$_{20}$ and PrV$_{2}$Al$_{20}$, respectively, as consistent with experimental observations. We note that the other subdominant contributions such as the diagonal components for $O_2^0$ and $O_2^2$ between the intersite $\mu=1$ and $\nu=2$ in the case of PrTi$_{2}$Al$_{20}$ (see Fig. 4 (c)) and the off-diagonal components between $O_2^0$ and $O_2^2$ at the same site in the case of PrV$_{2}$Al$_{20}$ (see Fig. 4 (d)) are also important to determine the ordered states below $T_{\rm Q}$ as shown lalter.

As the RKKY interactions obtained in the present study are long-range interactions between more than 62th nearest neighbor sites (see Figs. 3 (a)-(d)), the ordered states are expected to be described well with the mean-field approximation in contrast to the simplified models with the nearest neighbor interactions. Then, we solve the RKKY Hamiltonian eq. (\ref{Jrkky}) with the obtained RKKY interactions within the mean-field approximation and find that the FQ  and the AFQ orders take place in the cases of PrTi$_{2}$Al$_{20}$ and PrV$_{2}$Al$_{20}$, respectively, as shown in Figs. 5 (a) and (b), where the order parameters of the $i$-th unit cell are given by 
$\langle \hat X_\alpha^\mu(i) \rangle=\langle \hat X_\alpha^\mu \rangle $ 
for the FQ order and 
$\langle \hat X_\alpha^\mu(i) \rangle=\langle \hat X_\alpha^\mu \rangle e^{i\bm{Q}\cdot \bm{R}_i}$ with ${\bm{Q}}=(\pi/a,0,\pi/a)$ 
for the AFQ order. We have no octupole moment $\langle \hat T_{xyz}\rangle$ in the both cases of PrTi$_{2}$Al$_{20}$ and PrV$_{2}$Al$_{20}$ as consistent with the $\mu$SR experiments\cite{Ito2011,Ito2015}.

In the case of PrTi$_{2}$Al$_{20}$ shown in Fig. 5 (a), the order parameter changes its sign between the Pr site 1 and 2 in the same unit cell, that is due to the negative value of the site-off-diagonal components of the RKKY interactions at $\bm{q}=(0,0,0)$ as shown in Fig. 4 (c). The linear combinations of the obtained  order parameters  $\langle \hat O_2^0\rangle$ and $\langle \hat O_2^2\rangle$ yield the $O_2^0$-type FQ order with the principal $x$-axis as shown in Fig. 6 (a), where the order parameter changes its sign between the Pr site 1 and 2 in the same unit cell as mentioned above. The $O_2^0$-type FQ order is consistent with the neutron scattering and the ultrasonic experiments\cite{Sato2012,Koseki2011} but its sign is inconsistent with the NMR experiment\cite{Taniguchi2016} which revealed that the $O_2^0$ order parameters of the Pr site 1 and 2 have the same sign in contrast to the present result with the opposite sign. This discrepancy will be discussed later. The $T$-dependence of the order parameter seems to show the 2nd-order phase transition (see Fig. 5 (a)) as observed in experiments\cite{Sakai2011}, however, it actually exhibits the 1st-order phase transition with a tiny discontinuity $\sim 0.01$ which is hard to detect as consistent with the prediction from the Landau theory\cite{Taniguchi2019}. We also explicitly calculate the specific heat (not shown) and confirm that the calculated result well accounts for the observed one\cite{Sakai2011} except for a tiny latent heat.

\begin{figure}[t]
\centering
\vspace{+0.3cm}
\includegraphics[width=8.5cm]{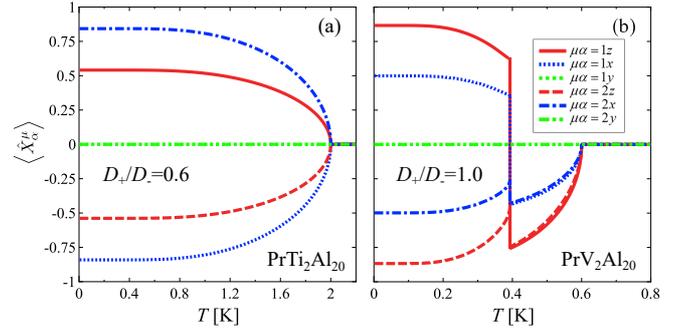}
\caption{(color online) 
Temperature dependence of the order parameters for the FQ order in the case of PrTi$_{2}$Al$_{20}$ (a) and that for the AFQ order  in the case of  PrV$_{2}$Al$_{20}$ (b).}
\label{Figure5}
\end{figure}

\begin{figure}[t]
\centering
\vspace{+0.3cm}
\includegraphics[width=7.0cm]{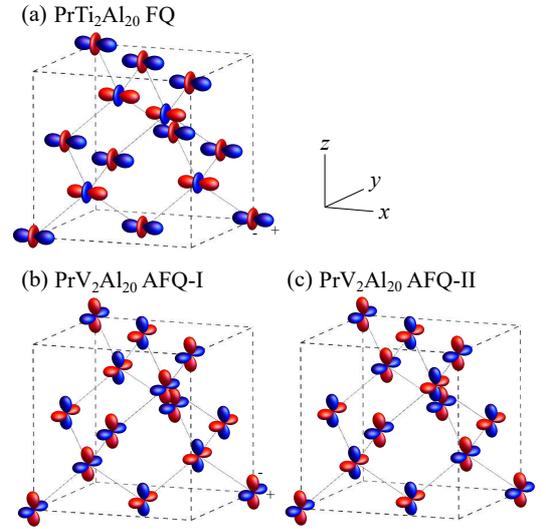}
\caption{(color online) 
Schematic configurations of the order parameters on the diamond lattice consist of Pr ions representing the $O_2^0$-type FQ order (a) in the case of PrTi$_{2}$Al$_{20}$ and the $O_2^2$-type AFQ orders, the high temperature AFQ-I (b) and the low temperature AFQ-II (c) in the case of PrV$_{2}$Al$_{20}$. $\pm$ indicates the sign of the charge distribution.}
\label{Figure6}
\end{figure}

In the case of PrV$_{2}$Al$_{20}$ shown in Fig. 5 (b), the system exhibits two distinct AFQ orders, the high temperature AFQ-I and the low temperature AFQ-II, and shows subsequent two phase transitions, the 2nd-order one from normal to the AFQ-I and the 1st-order one from the AFQ-I to the AFQ-II. This may be responsible for the double transitions observed by specific heat measurements\cite{Tsujimoto2015} although the 1st-order phase transition has not been detected. The order parameters of the AFQ-I are the same signs between the Pr site 1 and 2 in the same unit cell, that is due to the large positive value of the site-diagonal components of the RKKY interactions at the $\bm{q}=(\pi/a,0,\pi/a)$ as shown in Fig. 4 (d), while those of the AFQ-II are the opposite signs as shown in Fig. 5 (b). The linear combinations of the obtained  order parameters  $\langle \hat O_2^0\rangle$ and $\langle \hat O_2^2\rangle$ yield the $O_2^2$-type AFQ orders with the principal $x$-axis for the both AFQ-I and AFQ-II as shown in Fig. 6 (b) and (c), where the order parameters are the same (opposite) signs between the Pr site 1 and 2 in the same unit cell resulting in the stripe-like (staggered-like) AFQ order for AFQ-I (AFQ-II). The $O_2^2$-type AFQ order is observed in PrIr$_{2}$Zn$_{20}$\cite{Iwasa2017}, although the quadrupole moment of the AFQ order in  PrV$_{2}$Al$_{20}$ has not been determined by experiments so far.

\section{Summary and Discussions}
\label{section4}

In summary, we have investigated the multipole orders in PrTi$_{2}$Al$_{20}$ and PrV$_{2}$Al$_{20}$  on the basis of the RKKY interactions between the quadrupoles $\hat O_2^0$ and $\hat O_2^2$ and the octupole $\hat T_{xyz}$ of the Pr ions evaluated from the first-principles band calculations. What we have found are: In the case of PrTi$_{2}$Al$_{20}$, the 1st-order phase transition to the $O_2^0$ FQ order with a tiny discontinuity takes place. In the case of PrV$_{2}$Al$_{20}$,  the system exhibits two distinct $O_2^2$ AFQ orders, the high temperature stripe-like AFQ-I and the low temperature staggered-like AFQ-II, and shows subsequent two phase transitions, the 2nd-order one from the normal to the AFQ-I and the 1st-order one from the AFQ-I to the AFQ-II. The octupole moment is absent in the both cases. The obtained results seem to be consistent with the experimental observations.

Detailed analysis has revealed that the maximum in the RKKY interaction $J_{\alpha \beta }^{\mu \nu } (\bm{q})$ at ${\bm{q}}=(0,0,0)$ in the case of PrTi$_{2}$Al$_{20}$ is mainly due to the small circular hole FSs centered at the $\Gamma$ point. In the case of PrV$_{2}$Al$_{20}$, the number of electron relatively increases and the Fermi level shifts upward, resulting in the disappear of the small hole FSs and then the absence of the maximum in  $J_{\alpha \beta }^{\mu \nu } (\bm{q})$  at ${\bm{q}}=(0,0,0)$. This is nothing but the origin of the difference between the FQ order of PrTi$_{2}$Al$_{20}$ and the AFQ order of PrV$_{2}$Al$_{20}$. We have also performed the first-principles band calculations for La$T_{2}$Zn$_{20}$ ($T$=Ir, Rh) and have found that there are no small circular FSs similar to the case of PrV$_{2}$Al$_{20}$. This is consistent with the experiments where Pr$T_{2}$Zn$_{20}$ ($T$=Ir, Rh) as well as PrV$_{2}$Al$_{20}$ exhibit the AFQ orders\cite{Ishii2011,Ishii2013,Onimaru2010,Onimaru2012} while PrTi$_{2}$Al$_{20}$ exclusively exhibits the FQ order\cite{Koseki2011,Sakai2011} among the Pr 1-2-20 systems.

In the present study, we have considered the $c$-$f$ hybridization only between the $5d$ and $4f$ on the same Pr atom as the Pr $5d$ electrons give the dominant contribution to the susceptibility, and have neglected the inter-site $c$-$f$ hybridizations between the Pr-$4f$ and the other $c$-electrons such as the Al-$3p$. We have also exclusively considered the non-Kramers $\Gamma_{3}$ CEF ground states of the Pr ions with the $4f^2$ configuration and have neglected the CEF excited states such as the magnetic $\Gamma_{4}$ and  $\Gamma_{5}$ states which are considered to be important as the intermediate virtual states\cite{Hattori2014}. The discrepancy between the present study and the NMR experiment\cite{Taniguchi2016} concerning the sign of the order parameters between the Pr site 1 and 2 in PrTi$_{2}$Al$_{20}$ may be resolved by including those effects neglected here, but the essential feature of the FQ and the AFQ orders obtained from the present study are expected to be unchanged.

Below $T_{\rm Q}$, the superconductivity are found to coexist with the FQ and the AFQ orders for PrTi$_{2}$Al$_{20}$ and PrV$_{2}$Al$_{20}$, respectively\cite{Onimaru2010,Onimaru2012,Sakai2012,Tsujimoto2014} and are expected to be mediated by the quadrupole waves (quadrupolons) instead of the phonons in the BCS theory. Therefore, we need further calculations to obtain the quadrupole wave dispersions\cite{Shiina2003} responsible for the pairing interaction and also for various physical quantities such as the power-law behavior of the specific heat\cite{Tsujimoto2015}. Such calculations are now under way and explicit results will be reported in a subsequent paper.

\section*{Acknowledgments}
This work was partially supported by JSPS KAKENHI Grant Number JP 21K03399. Numerical calculations were performed in part using OFP at the CCS,University of Tsukuba and the MASAMUNE-IMR, Tohoku University. 

\bibliography{main.bib}

\begin{thebibliography}{10}
\bibitem{ratio}
When Ti is substituted by V in  Pr$T_2$Al$_{20}$ ($T$=Ti, V), the Fermi level 
$\varepsilon _{\rm F}$  shifts upward as shown in Figs. 2 (a) and (b) and then 
$D_+$ ($D_-$)  given in eq. (\ref{D+}) (eq. (\ref{D-})) increases (decreases) 
resulting in the increase in $D_+/D_-$. Therefore, $D_+/D_-$ for 
PrTi$_2$Al$_{20}$ is set to be smaller than that for PrV$_2$Al$_{20}$.
\end{thebibliography}
\bibliographystyle{jpsj.bst}

\end{document}